\documentstyle[preprint,aps]{revtex}
\begin{document}
\draft
\title
{\bf Accurate Four-Body Response Function with Full Final State Interaction:\\
Application to Electron Scattering off $^4$He 
}
\author{Victor D. Efros$^{1)}$, Winfried Leidemann$^{2)}$, and
Giuseppina Orlandini$^{2,3)}$}
\address{
1) Russian Research Centre, Kurchatov Institute, Kurchatov Square 1,
123182 Moscow, Russia\\
2) Dipartimento di Fisica, Universit\`a di Trento, 
 I-38050 Povo, Italy\\
3) Istituto Nazionale di Fisica Nucleare, Gruppo collegato di Trento
}

\date{\today}
\maketitle

\begin{abstract}
The longitudinal $(e,e')$ response function of
$^4$He is calculated precisely with full final
state interaction. The explicit calculation of the four-body continuum
states is avoided by the method of integral transforms. Precision tests
of the response show the high level of accuracy. Non--relativistic nuclear
dynamics are used. The agreement with experimental data is very good over
a large energy range for all considered momentum transfers ($q=300$, 400, 500
MeV/c). Only at higher $q$ the theoretical response overestimates the 
experimental one beyond the quasi-elastic peak.
\end{abstract}


A new method for the calculation of the inelastic response of an $N$-body system
to an external probe is proposed in Ref. \cite{elo94}.
It allows an exact calculation without the knowledge
of the $N$-body scattering state. The high level of accuracy of the method
has been shown for the longitudinal electron scattering responses of the
nuclear two- and three-body systems \cite{elo94,sara95}. The real superiority
of the approach, however, becomes evident when applied to a four-body system.
In fact a solution of the four-body medium energy continuum state problem
is presently out of reach, nonetheless four-body response functions can
be reliably calculated as pointed out in the following.
In this work we consider the important longitudinal electron scattering
response function $R_L$ of $^4$He which is calculated for the transfer
momenta $q$=300, 400, and 500 MeV/c. For $q=500$ MeV/c it is the first accurate
calculation with the final state nuclear interaction fully taken into
account. Our results are obtained within the framework of the
non--relativistic nuclear dynamics and using the single--particle form
of the electromagnetic operator. Such studies allow establishing the
limits of validity of this conventional  framework for the lightest
tightly bound nucleus.
Particularly interesting is the higher $q$ region. For more than a decade 
there has been a lot of discussion for complex nuclei regarding this 
region. An accurate calculation for $^4$He will help to shed some light 
in this range of $q$ values.

The idea of Ref. \cite{elo94} is to calculate the response in an indirect way.
First the Lorentz transform (LT)
\begin{equation}
\Phi(\sigma=\sigma_R+ i \sigma_I, q) =
\int d\omega {R(\omega, q)\over
(\omega-\sigma_R)^2 + \sigma_I^2}
\end{equation}
of the response function
\begin{equation}
R(\omega, q)=\sum_n|\langle n|
\Theta({\bf q})|0\rangle|^2\delta(\omega-E_n+E_0)
\end{equation}
is calculated, where $|0\rangle$ is the ground state of the system,
 $E_0$ is the ground state energy, $\Theta({\bf q})$ is the excitation
 operator, and
$\sigma_R>0$, $\sigma_I\neq 0$.
The solution of the following equation
\begin{equation}
(H-E_0-\sigma_R+ i \sigma_I) |\Psi\rangle = \Theta |0\rangle\,,
\end{equation}
leads directly to the LT:
\begin{equation}
\Phi(\sigma, q) = \langle \Psi | \Psi \rangle\,.
\end{equation}
In a second step  $R(\omega, q)$ is obtained via the
inversion of the transform.
The solution of Eq. (3) is unique. Indeed, the homogenous equation has only the
trivial solution because the hamiltonian $H$ has only real eigenvalues.
Since $\Psi$ has to fall off exponentially one
can use similar techniques as for the solution of the ground state problem.
Thus the extremely complicated asymptotic boundary condition of a four-body
scattering state has not to be considered at all.

In the past other integral transforms were proposed, namely
Stieltjes \cite{efros85} and Laplace transforms \cite{cs92,cs94}. 
The Laplace transforms of the longitudinal response were 
obtained with a realistic force for $q=300$ and 400 MeV/c via a
Green Function Monte-Carlo calculation (GFMC) \cite{cs92}.  
Also the Laplace transforms of the transverse response and the effects 
of two-body operators on the transforms in both longitudinal and 
transversal cases were considered  via a GFMC \cite{cs94}. Good agreement 
with the transforms of 
the experimental data is found. There is, however, a fundamental problem 
in obtaining response functions themselves from these 
transforms. Unlike the LT they 
sample contributions over a large energy range. This results in big problems 
for the inversion [6]. Nevertheless the longitudinal $R(\omega,q)$ of $^4$He 
has been obtained by an inversion of the Laplace transform for $q=400$ MeV/c 
\cite{cs92}. The result is rather similar to ours in Fig. 4. We are not able to 
fully interpret this agreement since the statistical errors of a 
GFMC lead to an uncertainty in the inversion of the 
Laplace transform. Unfortunately, the inversion error is not estimated in
ref. \cite{cs92}, which in general 
can be sizeable \cite{elo93}.  On the contrary, for the 
LT inversion problems are much less important 
\cite{elo94,sara95}. Moreover, the numerical effort for the calculation of the
LT seems to be much smaller than for the Laplace
transform. However, a fair comparison can only be made
when both calculations are performed for the same potential model.

Our nuclear Hamiltonian includes central even potentials\\
\begin{equation}
  V(ij)=V_{31}(r_{ij})P^+_{\sigma}(ij)P^-_{\tau}(ij)+V_{13}(r_{ij})
P^-_{\sigma}(ij)P^+_{\tau}(ij)
\end{equation}
providing realistic description of the $S$--wave phase shifts
up to the pion threshold. We construct the
$V_{31}$ and $V_{13}$ potentials by modifying the complete $NN$ interaction
of Ref. \cite{dtrs75}. The disregarded tensor force is effectively simulated
via a dispersive correction ($V\to V - V_{tensor}^2/const$). The potentials
obtained lead to almost the same phase shifts as in Ref. \cite{dtrs75}. A full
description of the potential will be published elsewhere \cite{elo96}.
It describes the static properties of $^4$He rather
well leading to a binding energy of 31.3 MeV and an rms radius of 1.40 fm.
Also the description of the elastic form factor is rather realistic up to its
first minimum. The present ansatz for the potential
will lead to results quite similar to those for more general nuclear forces.
The three--nucleon studies undertaken so far testify to this opinion
\cite{vmt92}.
Although more intensive the calculations with a completely realistic nuclear
force are also quite feasible within our approach.

In the following we describe the techniques we use
 for solving the dynamic equation (3).
We seek for the solution in the form of an expansion over the correlated
hyperspherical basis first used in Ref. \cite{fen72}. The expansion
converges quickly in few--nucleon bound state problems \cite{fen72,pisa}.
Our basis functions are of the form
\begin{equation}
J R_N(\rho)\left[Y^{[f]\mu}_{KLM}\left(\Omega\right)\theta^{[\bar{f}]
\bar{\mu}}_{S=0,T}\right]^a.
\label{eq:bas}
\end{equation}
Here $\rho$ is the hyperadius, $\rho =(\xi_1^2+\xi_2^2+\xi_3^2)^{1/2}$,
$\vec \xi_i$ are the normalized Jacobi vectors, 
and $\Omega$ denotes collectively
eight hyperangular variables. The quantities $Y^{[f]\mu}_{KLM}$ are
hyperspherical harmonics (HH) with
hyperangular $K$ and orbital $L,M$ momentum quantum numbers. These HH are
components $\mu$ of
irreducible representations $[f]$ of the four--particle permutation group
$S(4)$. The spin--isospin functions $\theta$ (see e.g. \cite{bad67}) enter
Eq. (\ref{eq:bas}) with the same spin and isospin values $S=0$, $T=0$, and
$T=1$ as in the expansion of the right--hand side of Eq. (3). They belong to
the conjugate representation
$[\bar{f}]$ of $S(4)$. The square brackets mean coupling to the function
antisymmetric with respect to permutations of both spatial and spin--isospin
particle coordinates. $R_N$ are the hyperradial functions, and
$J$ is the Jastrow correlation factor.

The system of equations for the expansion coefficients is obtained by
projecting Eq. (3) onto the subset of functions (\ref{eq:bas}) with $K$ up
to some $K_{max}$ and $N$ up to some $N_{max}$.
This system is split with respect to $L,M$ and $T$ values. Since $L(^4$He)$=0$
in our model the $L$ quantum number coincides with the multipole order of the
transition operator. The response, as well as Eq. (4), is independent of a
$\bf  q$ direction that can be chosen along
the $z$ axis. Only the $M=0$ value gives a non--zero contribution in this case.
The matrix elements are calculated with a Monte Carlo integration.

The HH entering Eq. (\ref{eq:bas}) are constructed by applying the convenient
form, see e.g. \cite{fom81}, of the Young operators to the simple
Zernike--Brinkman type
HH. The multiplicities of various $[f]$ representations at given $K$ and $L$
values are obtained as traces of the Young operators  calculated in
the Zernike--Brinkman basis \cite{fom81,dzh77}.

The hyperradial functions of the form \cite{fom81,ehr71} $R_N(\rho)
\sim L^8_N(\rho/b) \exp(-\rho/2b)$ are used.
Here  $L_N^{\alpha}$ are Laguerre polynomials, and $b$ is a scale parameter
which is kept the same for all the $\sigma$ values considered and is chosen
to  enable sufficiently fast overall convergence. The results are rather
insensitive to the $b$ values. The rate of the hyperradial convergence in our
case is lower than in the bound state calculations (e.g. \cite{fom81,ehr71}),
and better $R_N$ can perhaps be found.

The two-body correlation function $f(r)$ entering the Jastrow factor is taken
to be spin--independent and is chosen in a
conventional way. At $r\leq r_0$ it is a solution to the Schr\"odinger equation
with the potential taken as the half--sum of the triplet and singlet $NN$
forces. The $r_0$ point is chosen from the condition $f'(r_0)=0$.
At $r>r_0$ $f(r)=f(r_0)$. The kinetic energy matrix elements
with the Jastrow factor are cast to a convenient form \cite{fen72}.

We calculate the LT of $R_L(\omega,q)$
with $\sigma_I=20$ MeV. The quantities in Eqs. (2), (3) pertain to
the center of mass system, and
\[ \Theta({\bf q})= \sum_k \left[\frac{1+\tau_3(k)}{2}+
\frac{G_E^n(q^2)}{G_E^p(q^2)}\frac{1-\tau_3(k)}{2}\right]
e^{i{\bf q}({\bf r}_k-{\bf R}_{c.m.})} \]
where $G_E^{p,n}$ are nucleon Sachs form factors. In order to reach convergence
we choose a sufficiently large $N_{max}$ for the hyperradial functions
($N_{max}=20$, 25, and 30 for $q=300$, 400, and 500 MeV/c, respectively). 
The multipole transitions of the charge operator are
taken into account up to a maximal order $L_{max}$.
>From the evaluation of the various multipole contributions to the
Coulomb sum rule we find that the following $L_{max}$ values lead to
an exhaustion of the sum rule by more than 99\%:
$L_{max}=4$, 5, and 6 for $q=300$, 400, and 500 MeV/c, respectively.
These $L_{max}$ values are adopted at solving Eq. (3).
The maximal hyperangular order $K_{max}$ is taken equal to 7,
only in case of $L_{max}=6$ the value 8 is used. This is sufficient to
completely exhaust the various multipole strengths  for $q=300$ MeV/c.
Even for $q=500$ MeV/c one misses only a small fraction of the strength of
the less important multipoles with $L \ge 4$ (see also discussion below).

The results for the LT are shown in Fig. 1. Unlike
Stieltjes and Laplace transforms it is already obvious directly from the
LT that the response is governed by the quasi-elastic peak.
The inversion is performed with the same sets of basis
functions used in Refs. \cite{elo94,elo93}. Contrary to the nuclear two-
and three-body systems, we cannot of course compare the $R(\omega,q)$
obtained from the inversion with a direct calculation of the response
according to Eq. (2). Nonetheless it is possible to test the precision
of the response function results. A first test is the separate inversion of
all the various multipoles. It serves as a very important sum rule check,
since for a given multipole one can compare the sum rule from the
evaluation as ground state expectation value with that obtained from an
explicit integration of the response. This check leads to very good results
with relative errors of about 1\% for most of the transitions (average errors:
1.1\%, 1.0\%, and 2.0\% for $q=300$, 400, and 500 MeV/c, respectively).
Somewhat
larger errors are found only for $q=500$ MeV/c, where the less important
higher multipoles ($L \ge 4)$ are slightly underestimated by about 3\% -- 4\%.
As mentioned above  $K_{max}$ should be chosen somewhat larger for a complete
exhaustion of the strength of these multipoles.  Nonetheless we may say that
the sum rule results show the good accuracy of our method.
In Fig. 2 the isoscalar and isovector parts of the response function obtained
from the separate inversion are shown for $q=500$ MeV/c.
One sees that almost all multipoles have the typical structure due to the
quasi elastic peak. The only exception is the isoscalar Coulomb monopole
which exhibits a peak close to threshold. For the two lower momentum
transfers this C0 peak is
even more pronounced. For $q=300$ MeV/c its height reaches already one third
of the quasi-elastic peak height. The isovector strength is
twice as large as the isoscalar one.

Another very important check for the precision of the method
is obtained by the inversion of the total LT. The resulting
$R(\omega,q)$ should not differ from that obtained from
the separate inversion discussed above. 
Before discussing these results we should
mention that we encounter at low energy for $q=400$ and 500 MeV/c similar
inversion problems for the full LT as described in Ref. \cite{elo94}.
We solve this problem in a similar way as in Ref. \cite{elo94}, i.e. by
separate inversions for the sum of isoscalar C0 and C1 and for the sum
of all remaining multipoles; nevertheless in the following it will be called
total inversion.
The total response functions resulting from separate and total inversions
are shown in Fig. 3 for the three considered momentum transfers. From the
good agreement of the various curves it is evident that the inversion
is very unproblematic. Differences between the two inversion methods
are only found at lower energies, however they are quite unimportant.
We consider the inversion of the total $\Phi(\sigma,q)$ as the
more accurate result, since we obtain a better fit to the calculated LT 
in the low-energy region.
The total Coulomb sum rule is reproduced very precisely by the inversion
of the total LT. We find relative
errors of 0.2\%, 0.4\%, and 1.6\% for $q=300$, 400, and 500 MeV/c,
respectively. The reason for the somewhat larger error at $q=500$ MeV/c
has been already discussed above.

After having demonstrated the precision of the method we compare
our results with experimental data. To this end we have to consider
that the response function of Eq. (1) is defined for point particles. In order
to compare with experiment we have to multiply $R(\omega_{lab},q)$ with the
square of the proton charge form factor $G_E^p(q^2-\omega_{lab}^2)$,
where $\omega_{lab} = \omega + q^2/2M(^4$He). We take
the dipole fit to $G_E^p$ with the usual relativistic
correction \cite{def84}. In Fig. 4 we show our
results in comparison with experimental data \cite{bates,saclay}. It is
readily evident that for the lower $q$ value of 300 MeV/c the agreement
between theory and experiment is very good.  
The low--energy wings of the response at
$q$=400 and 500 MeV/c are also in a very good agreement with
experiment. In particular, the rather complicated threshold structure
of the experimental $R_L$ at $q=400$ and 500 MeV/c is described
extremely well.  
Beyond the quasi-elastic peak the theoretical result overestimates the
experimental one somewhat at $q=400$ MeV/c and in a more pronounced way
at $q=500$ MeV/c. If the experimental results are correct the theoretical 
formulation should include subnuclear and/or relativistic effects 
in order to remove the discrepancy.        
   
In conclusion we may say that we have successfully applied the method
of Ref. \cite{elo94} to a four-body system response to an external probe
with full final state interaction. This enabled us to calculate 
the accurate longitudinal response function of $^4$He. We have 
shown that the results are very precise. We obtain an excellent agreement 
with experiment at the momentum transfer of 300 MeV/c as well as for the 
low--energy wings at $q=400$ and 500 MeV/c.  At the latter $q$
values the theoretical results overestimate the experimental ones beyond
the quasi--elastic peak. Though somewhat more complicated a calculation
with a fully realistic potential model can also be carried out in a similar way.
The calculation of the transverse response with the present potential model 
is in progress \cite{elo96}.

The authors thank INFN for having provided a
dedicated work station (SUN SPARC-20) for the numerical calculations.
One of us (V.D.E.) thanks INFN for the financial support over the
period during which this work was carried out.

\vfill\eject

\begin{figure}
\caption{LT at $q$=300 (a), 400 (b), and 500 MeV/c (c)}
\end{figure}

\begin{figure}
\caption{Separate inversions of the various isoscalar (a) and
isovector (b) multipoles of the LT ($q$=500 MeV/c).
The various curves correspond to successive addition of multipole
contributions from $C0$ to $C6$}
\end{figure}

\begin{figure}
\caption{Response functions from total (full curves) and separate
inversions (dotted curves)}
\end{figure}

\begin{figure}
\caption{Response functions from total inversions with inclusion
of proton charge form factor (see text) in comparison
to experimental data}
\end{figure}


\begin{thebibliography}{99} \bibitem{elo94} V.D. Efros, W.
Leidemann, and G. Orlandini, Phys. Lett.  {\bf B 338}, 130 (1994).
\bibitem{sara95} S. Martinelli, H. Kamada, G. Orlandini, and W. Gl\"ockle,
Phys. Rev. {\bf C52}, 1778 (1995).
\bibitem{efros85} V.D. \'Efros, Sov. J. Nucl. Phys. {\bf 41}, 949 (1985).
\bibitem{cs92} J. Carlson and R. Schiavilla, Phys. Rev. Lett. {\bf 68}, 3682
(1992).
\bibitem{cs94} J. Carlson and R. Schiavilla, Phys. Rev. {\bf C49}, R2880
(1994).
\bibitem{elo93} V.D. Efros, W. Leidemann, and G. Orlandini, Few-Body Syst.
{\bf 14}, 151 (1993).
\bibitem{vmt92} E. van Meijgaard and J.A. Tjon, Phys. Rev. {\bf C45},
1463 (1992).
\bibitem{dtrs75} R. de Tourreil, B. Rouben, and D.W.L. Sprung, Nucl. Phys.
{\bf A242}, 445 (1975).
\bibitem{elo96} V.D. Efros, W. Leidemann, and G. Orlandini, paper in
preparation.
\bibitem{fen72} Yu. I. Fenin and V.D. \'Efros,
Sov. J. Nucl. Phys. {\bf 15}, 497 (1972).
\bibitem{pisa} A. Kievsky, M. Viviani and S. Rosati, Few--Body Sys.
Suppl. {\bf 7}, 278 (1994).
\bibitem{bad67} A.M. Badalyan, E.S. Gal'pern, V.N. Lyakhovitskii,
V.V. Pustovalov, Yu.A. Simonov, and E.L. Surkov,
Sov. J. Nucl. Phys. {\bf 6}, 345 (1968).
\bibitem{fom81} B.A. Fomin and V.D. \'Efros,
Sov. J. Nucl. Phys. {\bf 34}, 327 (1981).
\bibitem{dzh77} R.I. Dzhibuti, N.B. Krupennikova, and N.I. Shubitidze,
Theor. Math. Phys. (USSR) {\bf 32}, 704 (1977).
\bibitem{ehr71} G. Ehrens, J.L. Visshers, and R. Van Wageningen,
Ann. Phys. (N.Y.) {\bf 67}, 461 (1971).
\bibitem{def84} T. De Forest, Jr., Nucl. Phys. {\bf A414}, 347 (1984).
\bibitem{bates} S.A. Dytman $et$ $al.$, Phys. Rev. {\bf C34}, 800 (1988). 
\bibitem{saclay}  A. Zghiche $et$ $al.$, Nucl. Phys. {\bf A 572}, 513 (1994).
\end{thebibliography}
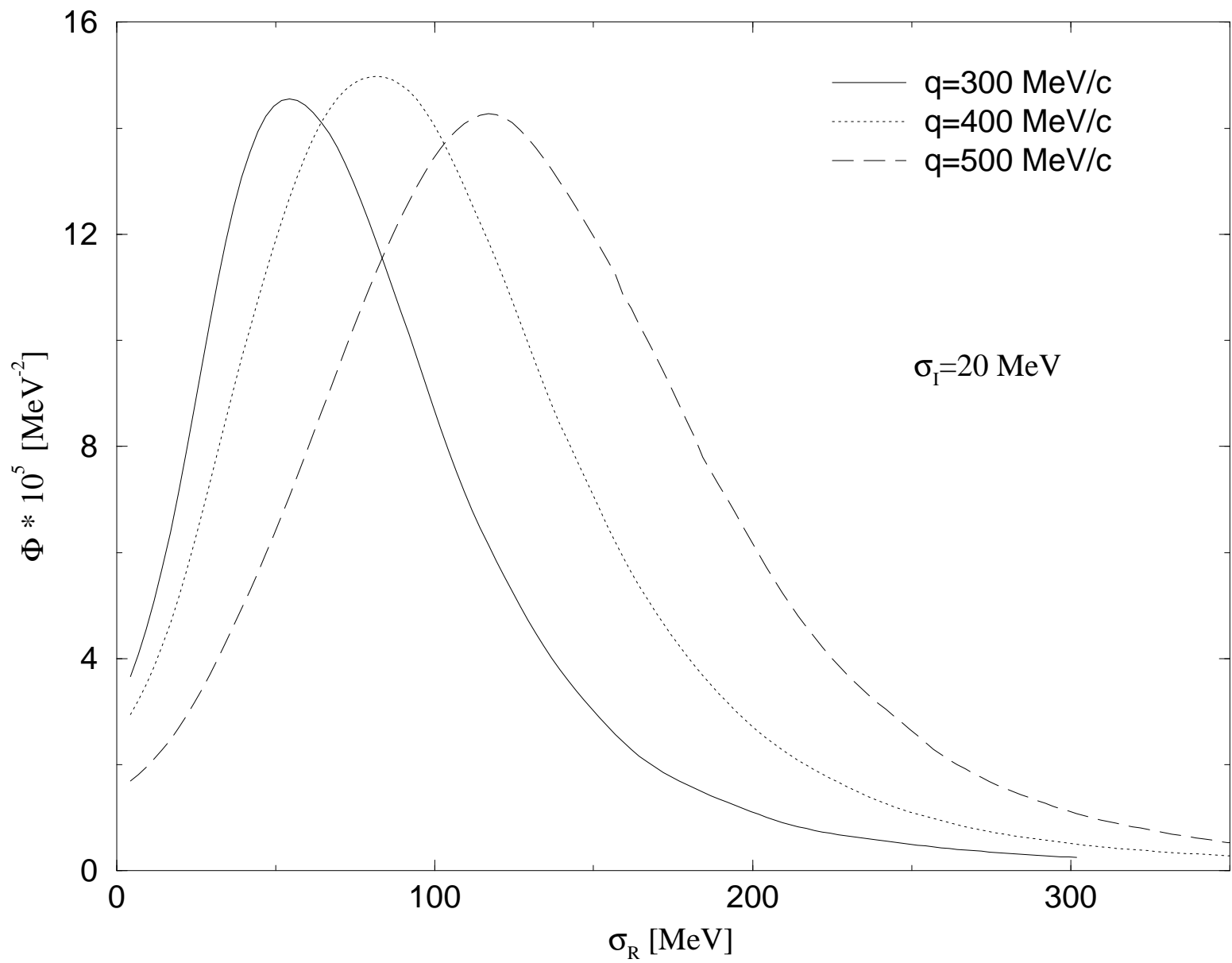
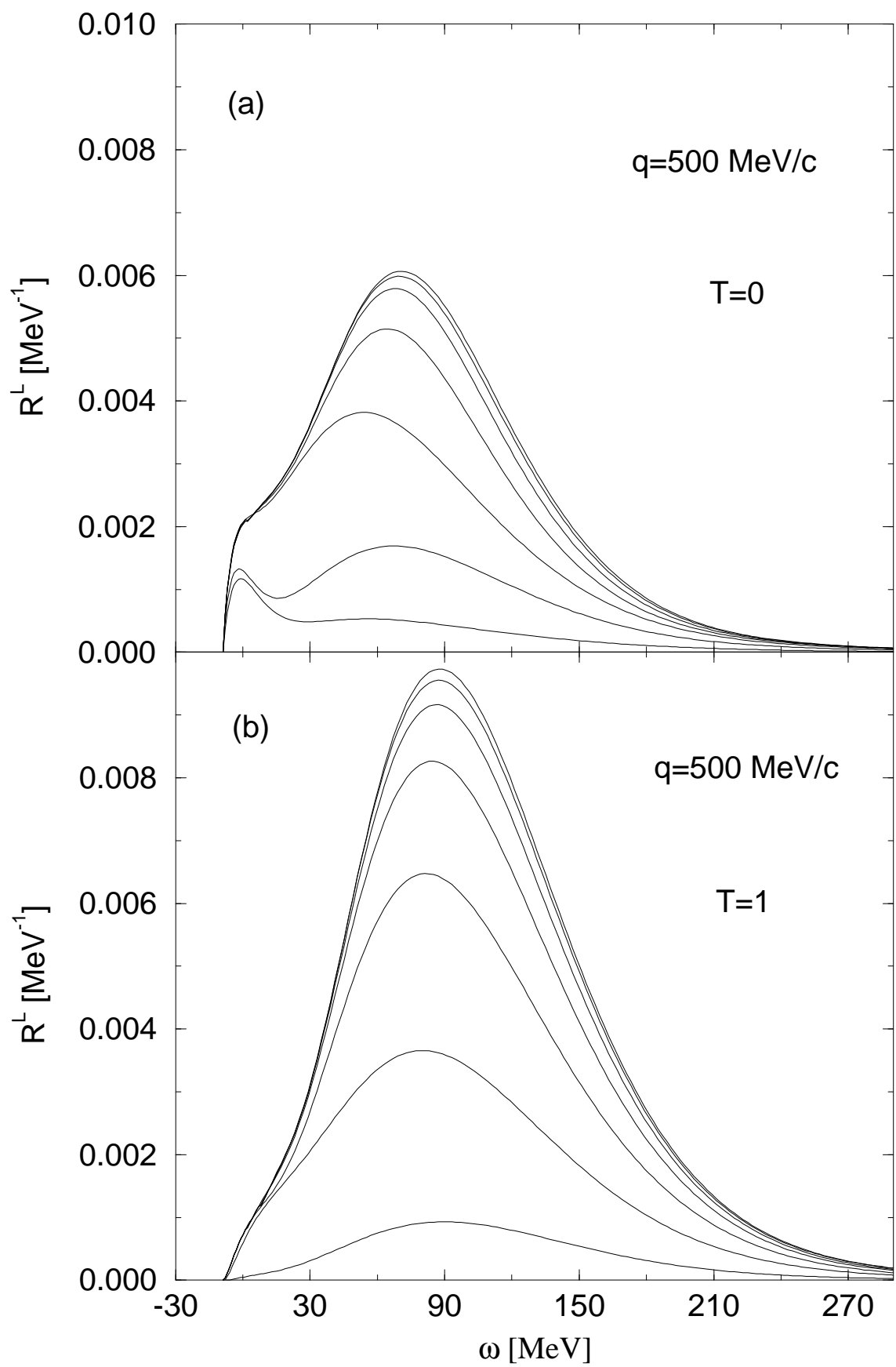
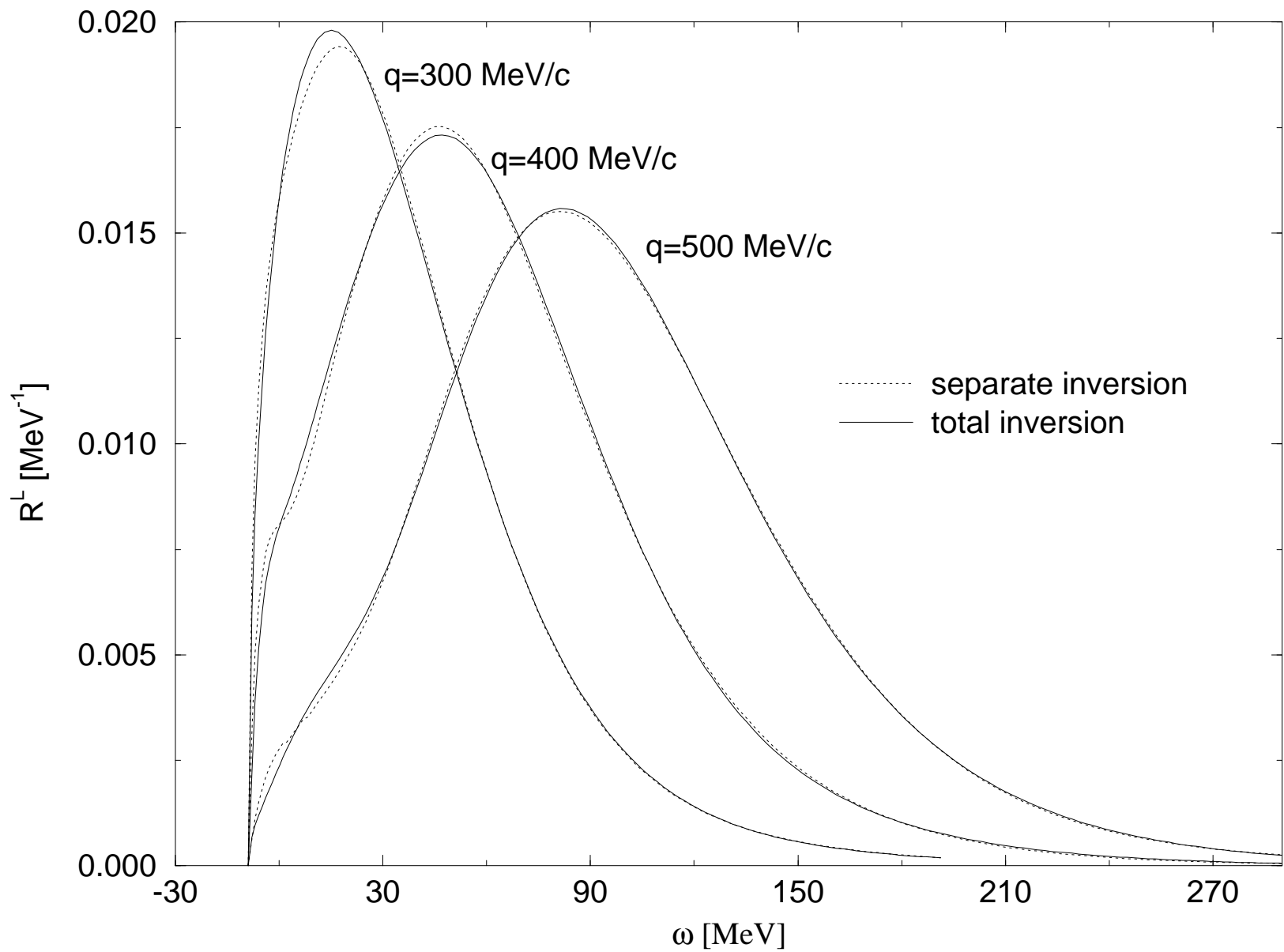
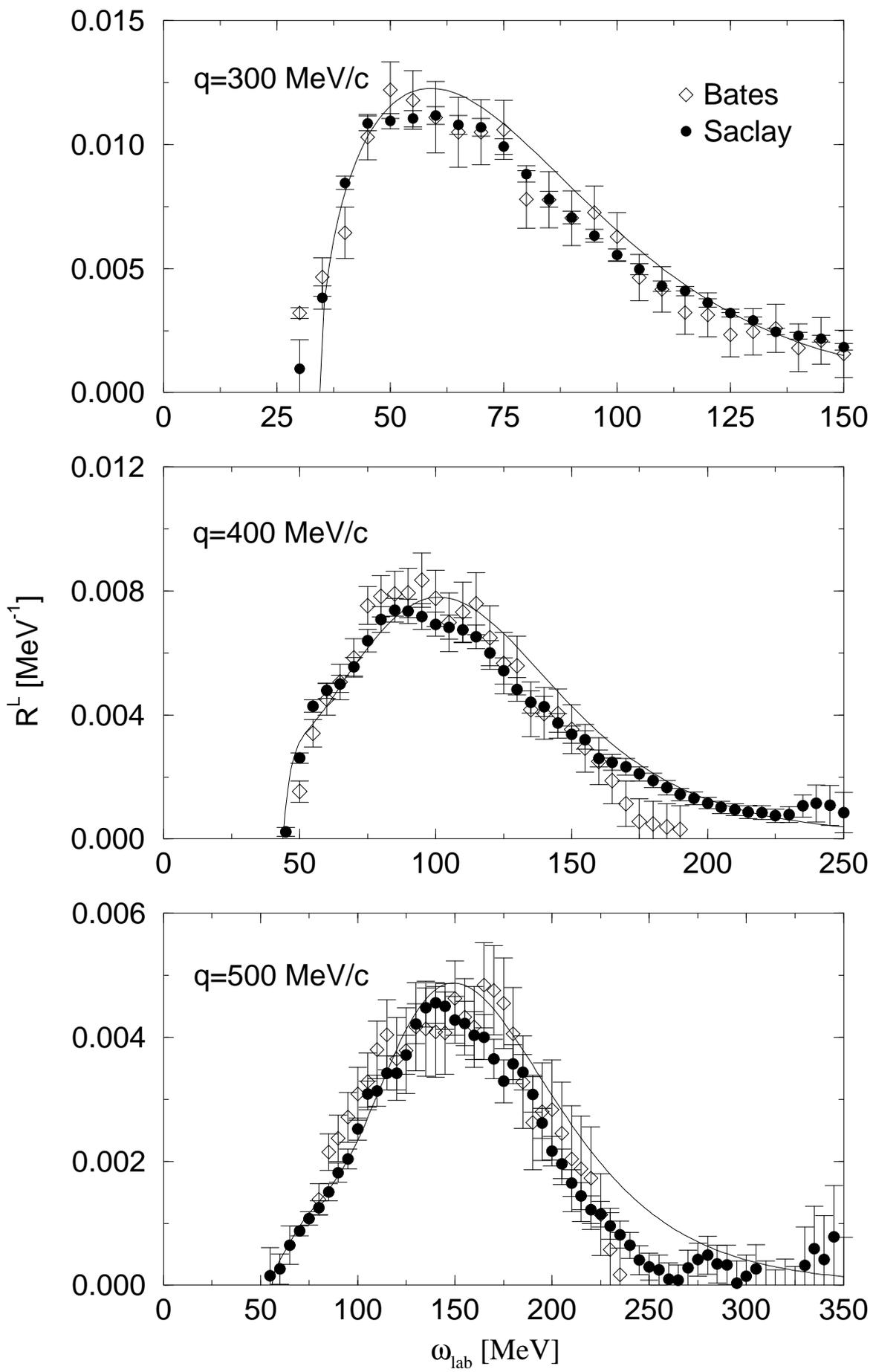
\end{document}